\documentstyle[preprint,aps,graphicx]{revtex}
\begin{document}
\draft
\title{\hspace{3.0in}{\small BGU PH-97/07}\\
\vspace{0.8cm}
Chiral Perturbation Theory Calculations For
S-wave $\pi^0$ production In pp Collision }
\author{ E.Gedalin\thanks{gedal@bgumail.bgu.ac.il},
 A.Moalem\thanks{moalem@bgumail.bgu.ac.il}
 and L.Razdolskaya\thanks{ljuba@bgumail.bgu.ac.il}}
\address{ Department of Physics, Ben Gurion University, 84105
Beer Sheva, Israel}
\maketitle
\begin{abstract}
The total cross section for the $pp \rightarrow pp \pi^0$ 
reaction at energies close to threshold is calculated within 
the frame of a chiral perturbation theory. The tree 
and one loop diagrams up to chiral order $D=2$ contributions are 
taken into account. The $L^{(0)}$ isoscalar-scalar part of 
t-channel two-pion exchange loop diagrams enhances the production 
amplitude strongly. The calculated cross section scale and energy 
dependence are very close to data.
\end{abstract}


\newpage

In two recent contributions\cite{park96,cohen96}, the cross 
section for the $pp \rightarrow pp \pi^0$ reaction at energies
 near threshold was calculated within the frame work of chiral 
 perturbation theory ($\chi$PT). Although $\chi$PT
accounts for all effects such as offshellness, unitarity and 
spontaneously broken chiral symmetry, the calculations of Refs.
\cite{park96,cohen96} underestimate the cross section 
data by a factor of 4-6. This stands in marked difference
with the results from traditional one-boson exchange (OBE) model
calculations, where contributions from heavy meson exchanges 
seem to resolve the discrepancy between predictions and 
data\cite{lee93,horowitz94,gedalin96}.Particularly, by applying 
a covariant OBE model the production amplitude is found to be 
dominated by the exchange of a $\sigma$-meson, due to a
strongly enhanced t-pole term\cite{gedalin96}.
It is the purpose of the present note to show that the 
failure of $\chi$PT calculations\cite{park96,cohen96}
to reproduce data may not be due to limitations 
of theory but to inconsistencies
in the way $\chi$PT was applied to this process and, that 
including loop contributions properly may
resolve the discrepancy between predictions and data.
  
A meson production in NN collisions necessarily involves 
large momentum transfers and, it is expected that loop diagrams 
describing short interactions may play an important role.
The exchange of an effective isoscalar-scalar $\sigma$-meson 
simulates the exchange of two correlated pions and, to the 
extent a t-pole term dominates the process, there must exist 
equally important contributions from the isoscalar-scalar part 
of t-channel two-pion exchange loop diagrams.
Clearly the existing $\chi$PT formalism\cite{park93},\cite{bernard94},
\cite{bernard95}already accounts for such contributions and there is no
need to introduce them externally. Basically, for the same reasons 
the low energy $\chi$PT coupling  constants encode only 
resonance meson exchanges which describe short range interactions of 
the Goldstone bosons and nucleons.

A prominent feature of a production process is that the transferred 
four momentum $q^2$ is space like and rather large. 
At threshold $q^2 =-Mm$, where $M$ and $m$ are the masses 
of the nucleon and the meson produced. This has two important 
consequences. The first is that off mass-shell
effects must be calculated exactly\cite{park96,cohen96} because 
these may enhance strongly the contributions from 
diagrams involving large momentum transfer. 
In practice, for the  $pp \rightarrow pp \pi^0$ reaction, the 
amplitude  corresponding to one-pion exchange including one-loop
contributions, hereafter to be noted by  $T(\pi^0 p \to \pi^0 p)$, 
becomes large due to offshellness.
The second consequence is concerned with the $\chi$PT 
expansion convergence.
The $\chi$PT expansion parameter is $q/M \sim q/(4\pi F)$. For 
low momentum transfer processes, $q \leq m$ the expansion parameter is
small and contributions from loop diagrams are rather small so that
their overall contributions to the transition amplitude is also small.
For the  $pp \rightarrow pp \pi^0$ reaction, however, the expansion 
parameter is relatively large\cite{cohen96}, of the order 
$\sim \sqrt {m/M}$, and each of the  Born,   
rescattering and one-loop diagrams terms is of the same
order of magnitude, {\it i.e.\/},  $m/M$.
Therefore, for the sake of consistency all three contributions
must be taken into account. Formally, the chiral order D=1 
Lagrangian coupling constants, $c_1$, $c_2$, $c_3$, are determined 
from $\pi$N elastic scattering data analysis which accounts for 
one-loop contributions, and using these constants to calculate 
the cross section for the  $pp \rightarrow pp \pi^0$  reaction 
would also require that these diagrams should not be neglected.

We follow Refs.\cite{gedalin96,moalem94,moalem95} and
write the transition operator for a production process, 
$NN \to NN B$, in the form,
\begin{equation}
{\hat T}_{23}={\hat Z}_{33} {\hat M}_{23}^{(in)} {\hat X}_{22}\ \ ,
\label{eq:2}
\end{equation}
where ${\hat M}_{23}^{(in)}$ is the primary production operator 
which describes the transition from a two-nucleon state  to a 
three-body state of two nucleon and a boson B, and ${\hat Z}_{33}$ 
and ${\hat X}_{22}$ are operators accounting for 
final state interactions (FSI) and initial state interactions (ISI) 
corrections, respectively.
The ${\hat M}_{23}^{(in)}$ contains all possible inelastic 
interactions while ${\hat Z}_{33}$ and ${\hat X}_{22}$ involve 
elastic interactions only between particles in the exit and entrance 
channels. From a diagrammatic point of view, the operator 
${\hat M}_{23}^{(in)}$ is the sum of all connected diagrams 
that begin with the two initial lines and end with
the three final particle lines. The correction terms ${\hat Z}_{33}$
and ${\hat X}_{22}$ are the sum of all diagrams (both connected and 
disconnected) that can be formed from elastic scattering blocks. 
It is to be noted that while ${\hat M}_{23}^{(in)}$ involves large 
momentum transfer ($q^2 \sim -M m$) the ISI and FSI correction 
factors involve small momentum transfers only. Accordingly,it is 
not possible to separate 
from the diagrams contributing to ${\hat M}_{23}^{(in)}$, 
sub-diagrams which involve initial or final particle elastic 
scattering only.

There exist, the so called reducible diagrams in the Weinberg's 
sense \cite{weinberg90}, with nucleons almost on the mass shell. 
These diagrams split into irreducible sub-diagrams by cutting 
such nucleon lines. Cohen et al.\cite{cohen96} apply a revised 
notion of irreducibility, where a sub-diagram is irreducible i it 
includes a small energy denominator $\sim m^2$ and irreducible 
otherwise. The reducible diagrams can be used to generate, by 
iterations, initial and final states wave functions. The 
$\chi$PT amplitude can then be derived as an overlap integral of 
these functions and the irreducible term ${\hat M}_{23}^{(in)}$. 
The kinematical conditions o a production process are such that 
each o the diagrams of ${\hat M}_{23}^{(in)}$ contains at least 
one connected sub-diagram with large momentum transfer. So that 
the factorization o the amplitude, Eq.\ref {eq:2}, is equivalent 
to the $\chi$PT transition amplitude of Cohen et al.\cite{cohen96}.

Now we turn to calculate the transition amplitude for
the $pp \rightarrow pp \pi^0$ reaction.
We use the $\chi$PT  pion-nucleon sector heavy-fermion formalism 
(HFF) Lagrangian of the form, 
\begin{equation}
L = L^{(0)}\ +\ L^{(1)}\ +\ L^{(2)}\ \ ,
\label{eq:2}
\end{equation}
where,
\begin{eqnarray}
&L^{(0)} = \frac {1}{2} [ (\partial_{\mu} {\bf \pi})^2 
- m^2 {\bf \pi}^2 ] - \frac{1}{6F^2} [ {\bf \pi }^2 
(\partial_{\mu} {\bf \pi})^2 - ({\bf \pi }
\dot \partial_{\mu} {\bf \pi})^2 ] + N^{\dagger}(v\partial)N 
+\\ \nonumber
& N^{\dagger}[ -\frac {1}{4F^2} {\bf \tau \dot \pi} \times 
(v\partial) {\bf \pi}
 - \frac {g_A}{F} S^{\mu} {bf \tau } [\partial_{\mu}
{bf \pi} + \frac {1}{6F^2}({\bf \pi \pi} \partial_\mu {\bf \pi} - 
\partial_{\mu}{\bf \pi \pi}^2 ] N;\\
&L^{(1)} = \frac {1}{2M} (v^{\mu} v^{\nu} - g^{\mu \nu})
[ N^{\dagger}\partial_{\mu} \partial_{\nu} ]N +
\frac {1}{4F^2} [iN^{\dagger} {\bf \tau \pi} \times \partial_{\mu}
{\bf \pi} \partial_{\nu}N + h.c.]\\ \nonumber
& + \frac {1}{2MF^2} N^{\dagger} [(c'_2 -
\frac {g^2_A}{4})(v\partial {\bf \pi})^2 - 
c'_3 (\partial_{\mu}{\bf \pi})^2 
- 2c'_1 m^2 {\bf \pi}^2] +...\ \ ,
\label{eq.3}
\end{eqnarray}
and 
\begin{equation}
L^{(2)} = - \frac {d_1}{MF}[i N^{\dagger} {\bf \tau } 
(v \partial{\bf \pi})
 S^{\mu} \partial_{\mu}N N^{\dagger}N + h.c.] + ...\ \ .
\end{equation}
Here $\pi$ and $N$ are the pion and nucleon fields, 
$v^{\mu}$ is the nucleon four velocity, $(v\partial)=
v^{\mu}\partial_{\mu}$
and, $c'_1$, $c'_2$ and $c'_3$ denote the dimensionless 
low energy 
coupling constants. These are determined from 
fitting S-wave $\pi^0$N scattering data to be\cite{bernard95},
$c'_1 = -1.63$, $ c'_2 = 6.20$ and $ c'_3 = - 9.86$.
To obtain the $\pi$ production amplitude to order $D=2$ one has 
to include, in addition to the tree diagrams, one loop diagrams 
with vertices from $L^{(0)}$. The main contributions to $M^{(in)}$ 
are due to the diagrams
of Fig. 1. The graphs 1a and 1b are the usual impulse and 
rescattering diagrams. Both are taken into account by 
Park et al.\cite{park96} and Cohen et al.\cite{cohen96}. 
The two graphs 1c and 1d which correspond
to two-pion t-channel exchange having 
isoscalar-scalar quantum numbers, yield a substantial
contribution to the amplitude. This is an analogous 
contribution to that from the t-pole $\sigma$-meson exchange 
(Fig. 2d and its cross counterpart) of the OBE model of 
Ref.\cite{gedalin96}. Albeit, the sum of 
graphs 1a-1d is equivalent to the off mass-shell $T(\pi^0 p 
\rightarrow \pi^0 p)$ amplitude (Fig. 2a).

The other graphs 1e and 1f are contributions specific to the 
production process only and can not be reduced to one-pion exchanges. 
In Fig. 1e, the quantum numbers in the $N{\bar N} \rightarrow 
N{\bar N} \pi$ channel correspond to an isoscalar-scalar state. 
Like the graphs in Figs. 1c-1d, this diagram simulates the
effective $\sigma$-meson exchange of the OBE model but with 
short range $\sigma p \rightarrow \pi^0 p$ production amplitude. 
The graph 1f is a short-range interaction mechanism provided by 
$L^{(2)}$.

The sum of all contributions from the graphs of Fig. 1 can now be 
written as,
\begin{equation}
M^{(in)}(pp \to pp\pi^0) = M^{(1)}_P + M^{(1)}_R + M^{(1)}_L 
+M^{(2)}_L + M^{(2)}_S.\label{mattot}
\end{equation}
where, 
\begin{eqnarray}
&M^{(1)}_P = -i \frac {g^3_A}{8MF^3(Q^2 - m^2)} {\bf Q}^2{\bf p}
{\bf \sigma }_1 + [1 \leftrightarrow 2, 3 \leftrightarrow 4], \\
&M^{(1)}_R = i \frac{g_A}{2MF^3(Q^2 - m^2)} [(c'_2 + c'_3 
- \frac{g^2_A}{4}) mQ^0 -2c'_1 m^2]{\bf p}{\bf \sigma }_1 
+ [1 \leftrightarrow 2,3 \leftrightarrow 4];\\
&M^{(1)}_L = i \frac {g^3_A}{12F^5 (Q^2 - m^2)} [6mQ^0 - 2Q^2 
-\frac{5}{2}m^2] B(q^2) {\bf p}{\bf  \sigma }_1 
+ [1 \leftrightarrow 2,3 \leftrightarrow 4]; \\
&M^{(2)}_L = i \frac{g^3_A}{24F^5} B(q^2){\bf p}{\bf \sigma }_1 + 
[1 \leftrightarrow 2, 3 \leftrightarrow 4];\\
&M^{(2)}_S = i\frac {d_1 \omega }{2FM} {\bf p}{\bf \sigma }_1 + 
[1 \leftrightarrow 2, 3 \leftrightarrow 4]\ \ ,
\label{mats}
\end{eqnarray}
where $\sigma$ denote the usual Pauli matrices, 
F and $g_A$ are the pion radiative decay and axial vector
coupling constants. 
The quantities $M_P^{(1)}$, $M_R^{(1)}$, $M_L^{(1)}$
denote one-pion exchange pole, rescattering and one-loop 
contributions, and $M_L^{(2)}$ and $M_S^{(2)}$ are contributions 
from graphs 2e and 2f. In Eqns.\ref{mattot}-\ref{mats}, 
$p=(E=M+{\bf p}^2/2M, {\bf p})$ and $k=(\sqrt {m^2+{\bf k}^2/2M},
 {\bf k})$ stand for the center of mass
(CM) four momenta of the incoming proton and pion produced; 
$Q=(Q^0={\bf p}^2/2M, {\bf p})$ and 
$q=({\bf p}^2/2M, {\bf p})$ are the transferred four momenta.
The bracket $[1 \leftrightarrow 2, 3 \leftrightarrow 4]$ 
represents the contributions from the same diagram with 
the proton momenta $p_1,\ p_3$ interchanged with $p_2,\ p_4$, 
respectively. The loop function is given by\cite{bernard95},
\begin{eqnarray}
B(q^2) = (-3 + 2 {\bf p^2} \frac{d}{dq^2}) B_0(q^2),\\
B_0(q^2) = -\frac{1}{16 \pi} \int_0^1 dz  \sqrt{m^2 - q^2 z(1 - z)}.
\end{eqnarray}

In the calculations to be presented below the values of constants 
and masses are taken to be: $F=93. MeV$, $m=135 MeV$, $M=938 MeV$ 
and $g_A =1.26$. There remains only one unknown constant, $d_1$ of 
Eqn. 5, to be  determined. To do this we follow a similar
procedure to that used in Refs.\cite{bernard95,cohen96}, i.e,
assuming that the short-range interactions originate from 
$\rho$  and $\omega$ vector meson exchanges
only, and calculate these applying the covariant OBE model of 
Ref.\cite{gedalin96}. This procedure yields,
\begin{equation}
d_1 =  \frac {f_{\pi NN} F}{ M}\left( \frac {g^2_{\rho NN}
 (1 + \kappa)}{m^2_{\rho} + Mm} + \frac {g^2_{\omega NN}}
 {m^2_{\omega}+ Mm}\right ) \ \ .
\end{equation}
Here $m_{\rho} = 770 MeV$, $m_{\omega} = 782 MeV$ are the masses 
of the $\rho$  and $\omega$ mesons, $f_{\pi NN}$ the $\pi NN$ 
pseudovector coupling
constants, $g_{\rho NN}$ and $g_{\omega NN}$ the $\rho NN$ and 
$\omega NN$ vector coupling constants and $\kappa$ is the ratio 
of tensor to vector $\rho NN$ coupling constants. Their values 
are taken from Machleidt et al.\cite{machleidt87}.

Before presenting numerical results we make two remarks. The first 
is that the $\pi ^0 p \to \pi^0 p$ elastic scattering amplitude is 
enhanced significantly because of offshellness and that the one-loop
term $M_L^{(1)}$  depends strongly on the mass ($Q^2$)
of the pion exchanged. At $Q= (m/2, \sqrt {Mm})$ it becomes 
sufficiently large  so that the sum of $M_R^{(1)} + M_L^{(1)}$, 
although cancels in part by the $M_P^{(1)}$ term, can still make an 
important contribution to the $M^{(in)}(pp \rightarrow pp \pi^0)$.
The second remark is concerned with the large momentum transfer through
one-loop diagrams. At least one of the pion lines in a loop  
must have a large momentum ( $\sim \sqrt {M m}$ ) and therefore, 
this loop represents a mechanism which takes place over short 
distances. Since the loop terms dominate the production amplitude, 
then the $pp \rightarrow pp \pi^0$ reaction occurs mainly 
over short distances in full agreement with the results
from OBE calculations\cite{gedalin96}. In fact there is a close
correspondence between the different $\chi$PT  contributions and 
those of the OBE model. 
First, the overall contributions from the impulse and rescattering 
terms (diagrams 1a and 1b) is rather small\cite{park96,cohen96} as 
suggested by OBE model\cite{gedalin96} for the diagram 2a. Secondly, 
the main contribution to the production amplitude is due to the 
diagrams 1c and 1d which 
are the equivalent to contribution from diagram 2d. 
Thirdly, the contribution of the graph 2c and its cross diagram to
the OBE model production amplitude are as negligibly small as the 
corresponding one-loop $\chi$PT  diagrams (not shown in Fig. 1). 
However, the OBE model calculations\cite{gedalin96} do not account 
for the short range part of the $\sigma p \to \pi^0 p$ amplitude 
corresponding to the contribution of diagram 1e.

Our predictions for the cross section are presented in Figs. 3-4 along
with the data of Refs.\cite{bondar95,meyer92}. All curves are 
corrected for FSI following the procedure described in 
Ref.\cite{gedalin96}. The dash-dotted curve displays our results with 
all of the terms of Eqn. 6  included and agrees rather well with 
the results of our OBE model calculations \cite{gedalin96} 
(given by the solid line).
The curve $\chi$PT2 gives the predictions without the short-range 
interaction term $M_S^{(2)}$ taken into account. This term interferes 
 destructively with the loop terms and brings the calculated cross 
 section to close agreement with data. The curve $\chi$PT1 
 represents the results of Cohen et al.\cite{cohen96}. We may thus 
 conclude that taking into account the contributions from one-loop 
 diagrams to chiral order D=2, $\chi$PT reproduces the cross section of 
 the $pp \rightarrow pp \pi^0$ reaction. With one-loop
contributions, there is a close correspondence between $\chi$PT  and 
the OBE model.

The calculations above can be improved by first including small 
$\sim 1/m$ correction terms, arising in HFF. The transferred 
momentum in the the $pp \rightarrow pp \pi^0$ reaction is rather 
large and the convergence of the HFF expansion is still to be 
verified. Secondly, contributions from other degrees of freedom 
must be taken into account. Excitations from the $\Delta$ (1232 MeV) 
nucleon isobar may well contribute to any of the 
graphs Fig. 1a-1e. Also, the vector meson contributions can be 
included explicitly, perhaps by using more generalized $\chi$PT. 
To go beyond that to allow for $\eta$ -meson excitations would 
require a broken SU(3) symmetry flavor version of $\chi$PT. 
The $\eta$ -meson is a Goldstone boson and hence
it must be treated on the same footing as the pion. Finally, 
as we have already indicated, the contribution from D=2 loop 
diagrams has the same order of magnitude as  those from the lower 
order terms. The contributions from the next D=3 chiral order
diagrams must then be calculated in order to ascertain the 
convergence of the  $\chi$PT  expansion.
Rough estimates indicate that such contributions may modify the 
calculated cross section by a few tens of per cents.

\vspace{1.5 cm}
{\bf Acknowledgments} This work was supported in part 
by the Israel Ministry Of Absorption.
We are indebted to  Z. Melamed
for assistance in computation.

\newpage
{\bf Figure captions}
\bigskip

\begin{figure}
\caption{The $\chi$PT diagrams giving the main contribution to the 
primary production amplitude for the 
$NN \rightarrow NN \pi^0$ reaction.}
\end{figure}

\begin{figure}
\caption{The OBE model diagrams contributing to the ${\hat 
M}_{23}^{(in)}$ for the $pp \to pp \pi^0$ reaction.}
\end{figure}

\begin{figure}
\caption{Predictions for the total cross section of the 
  $pp \rightarrow pp \pi^0$  reaction  vs. $Q_{cm}$, 
  the energy available in the center of mass system. 
The curve $\chi$PT2 gives the predictions without the 
short-range interaction term and the dash-dotted curve 
displays our results with all contributions included.
The curve $\chi$PT1 represents the results of Cohen et 
al.\protect\cite{cohen96}. Our OBE model predictions\protect\cite{gedalin96} 
are drawn as a solid curve. The data points are taken 
from Refs.\protect\cite{bondar95,meyer92}.}
\end{figure}

\begin{figure}
\caption{Predictions for the total cross section 
 vs. $\eta_{max}$, the maximal available 
momentum of pion in the center of mass system.}
\end{figure}

\newpage
\includegraphics[scale=0.75]{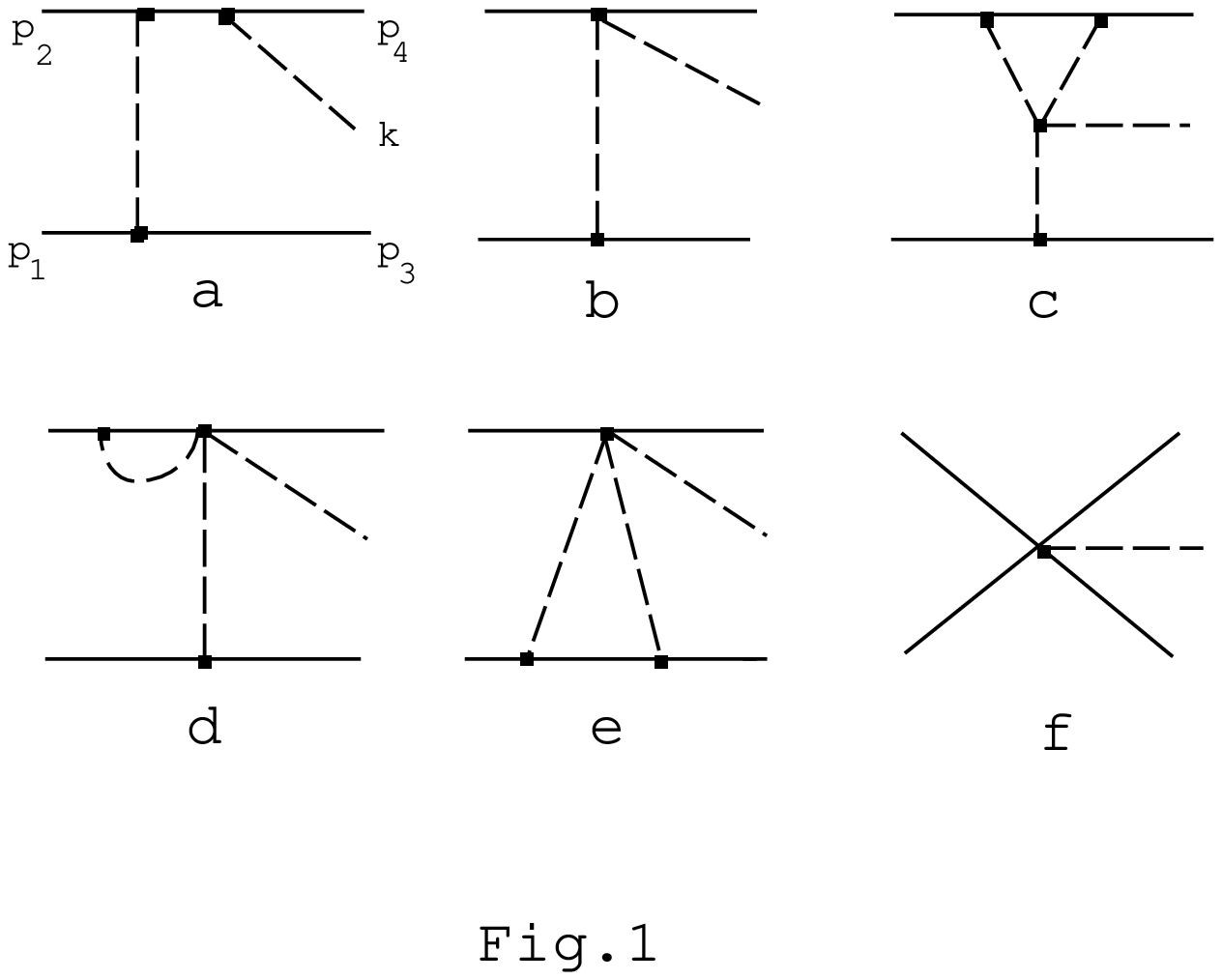}

\newpage
\includegraphics[scale=0.75]{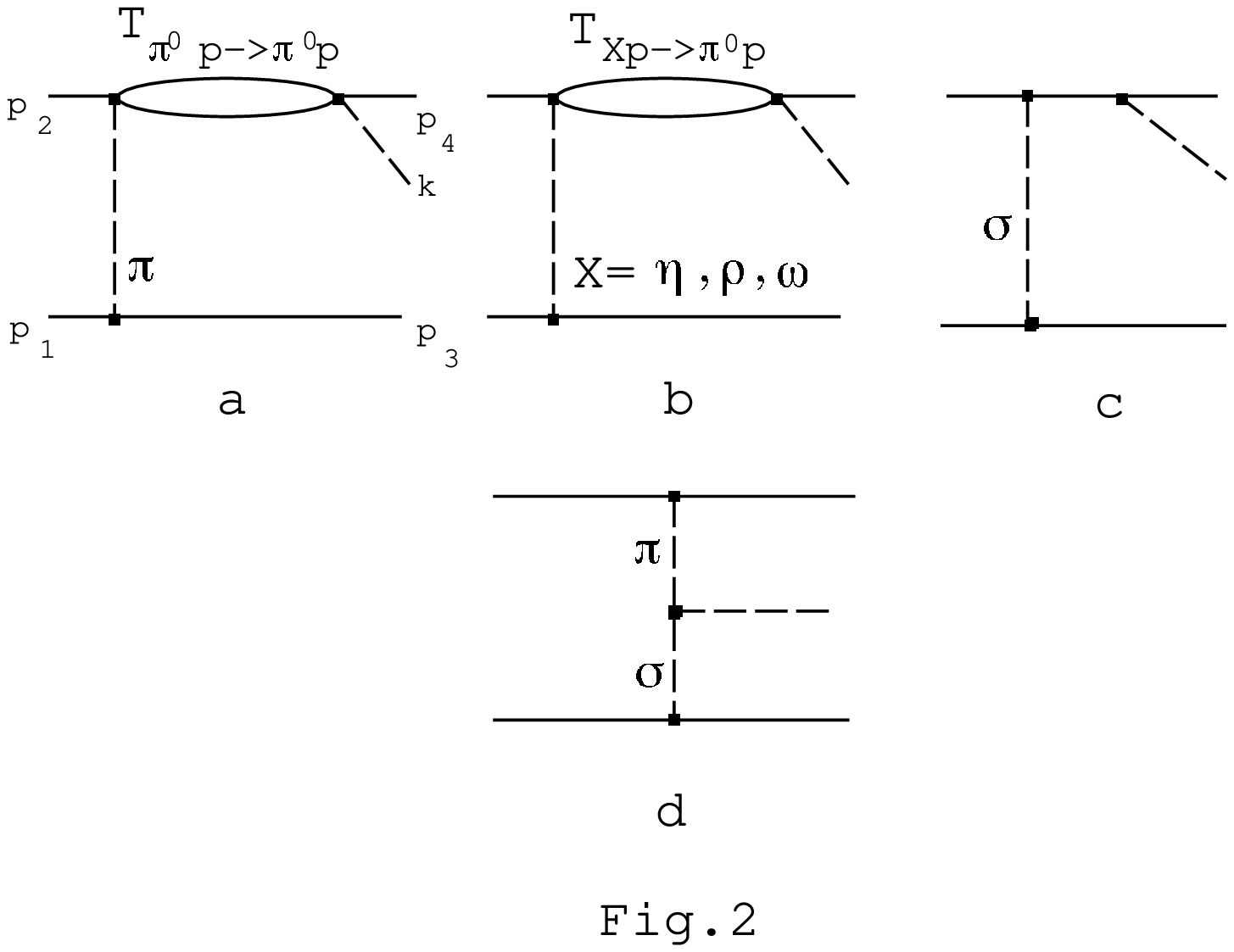}

\newpage
\includegraphics[scale=0.75]{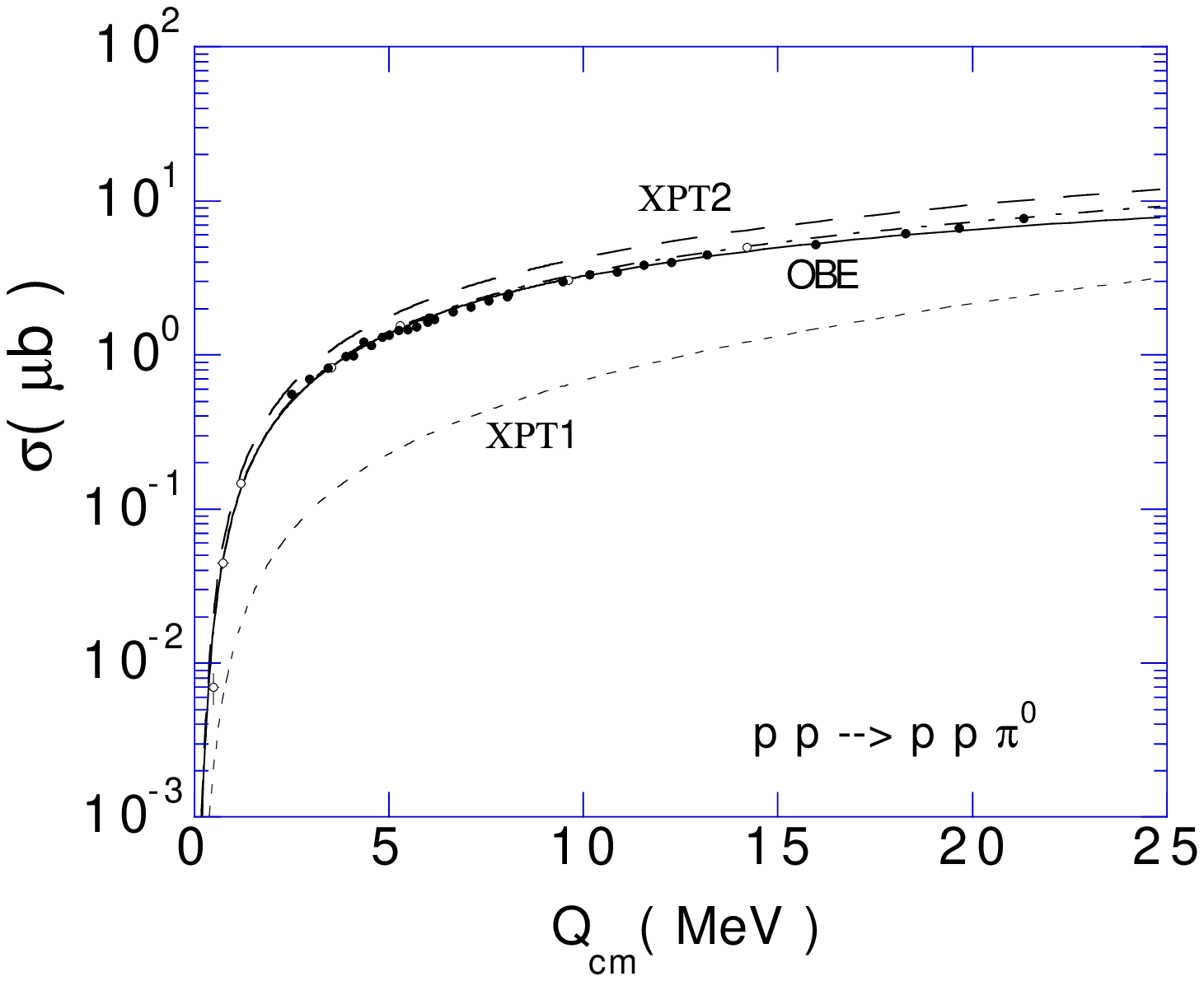} 

\newpage
\includegraphics[scale=0.75]{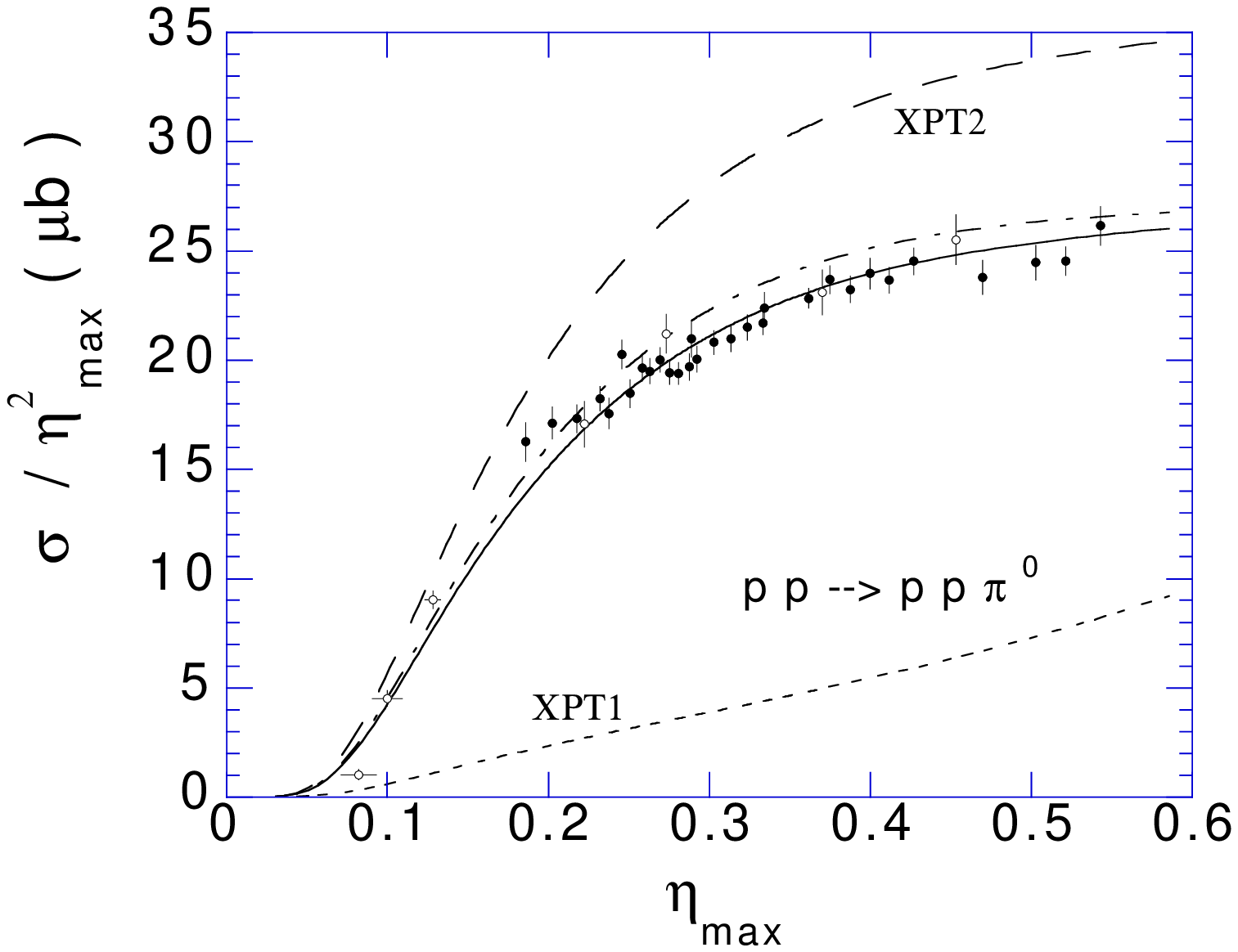}


\begin{thebibliography}{99}
\bibitem {park96} B.-Y.Park et al., Phys. Rev.
{\bf C53} (1996) 1519.
\bibitem {cohen96} T.D.Cohen et al., Phys. Rev. ${\bf C 53}$ 
(1996) 2661. 
\bibitem{lee93} T.S.H.Lee and D.Riska, Phys. Rev. Lett. 
${\bf 70}$(1993) 2237.
\bibitem {horowitz94} C.J.Horowitz et al., Phys. Rev.
{\bf C49} (1994) 1337.
\bibitem {gedalin96} E.Gedalin, A.Moalem and L.Razdolskaja, 
nucl-th/9611005.
\bibitem{park93} T.S.Park, D.-P.Min and M.Rho,   Phys. Rep.
	 ${\bf 233}$(1993) 341.
\bibitem{bernard94} V.Bernard, N.Kaiser, T.-S.H.Lee, and  
	Ulf-G.Meissner,  Phys. Rep. ${\bf 246}$(1994) 315.
\bibitem {bernard95} V.Bernard, N.Kaiser and Ulf-G.Meissner,
Int. J. Mod. Phys. ${\bf E4}$ (1995) 193.	
\bibitem {moalem94} A.Moalem, L.Razdolskaja and E.Gedalin, 
HEP-PH/9505264; \\
A. Moalem, E.Gedalin, L.Razdolskaja and Z.Shorer,
 $\pi$ N Newsletter Proceeding 
of the 6th International Symposium on Meson-Nucleon Physics and 
the Structure of the Nucleon, ${\bf 10}$ (1995) 172. 
\bibitem {moalem95} A.Moalem, E.Gedalin, L.Razdolskaja and 
Z. Shorer, Nucl. Phys. ${\bf A589}$ (1995) 649; \\
A. Moalem, E.Gedalin, L.Razdolskaja and Z.Shorer,
 Nucl. Phys.  ${\bf A600}$ (1996) 445.
\bibitem {weinberg90} S.Weinberg, Phys. Lett. ${\bf B251}$ (1990) 288;
$ibid$ ${\bf B295}$ (1992) 114.
\bibitem {machleidt87} R.Machleidt et al., Physics Reports ${\bf 149}$ 
(1987) 1.
\bibitem {bondar95} A.Bondar et al., Phys. Lett. ${\bf B 356}$ (1995) 8.
\bibitem {meyer92} H.O.Meyer et al., Nucl. Phys. ${\bf A539}$ (1992) 683.
\end{thebibliography}
\end{document}